# Superconducting proximity effect in a topological insulator using Fe(Te,Se)

He Zhao[1], Bryan Rachmilowitz[1], Zheng Ren[1], Ruobin Han[1], J. Schneeloch[2], Ruidan Zhong[2], Genda Gu[2], Ziqiang Wang[1] and Ilija Zeljkovic[1,¶]

[1]Department of Physics, Boston College, 140 Commonwealth Ave, Chestnut Hill, MA 02467;  [2]Brookhaven National Laboratory, Upton, NY 11973; ¶Corresponding author ilija.zeljkovic@bc.edu

## Abstract

Interest in the superconducting proximity effect has recently been reignited by theoretical predictions that it could be used to achieve topological superconductivity. Low-$T_c$ superconductors have predominantly been used in this effort, but small energy scales of ~1 meV have hindered the characterization of the emergent electronic phase, limiting it to extremely low temperatures. In this work, we use molecular beam epitaxy to grow topological insulator $Bi_2Te_3$ in a range of thicknesses on top of a high-$T_c$ superconductor Fe(Te,Se). Using scanning tunneling microscopy and spectroscopy, we detect $\Delta_{ind}$ as high as ~3.5 meV, which is the largest reported gap induced by proximity to an *s*-wave superconductor to-date. We find that $\Delta_{ind}$ decays with $Bi_2Te_3$ thickness, but remains finite even after the topological surface states had been formed. Finally, by imaging the scattering and interference of surface state electrons, we provide a microscopic visualization of the fully gaped $Bi_2Te_3$ surface state due to Cooper pairing. Our results establish Fe-based high-$T_c$ superconductors as a promising new platform for realizing high-$T_c$ topological superconductivity.

When a superconductor (SC) is brought into contact with a non-superconducting (normal) material, Cooper pairs from the SC can diffuse into the normal material, which in turn becomes superconducting. This phenomenon, known as the superconducting proximity effect (SPE), has been discovered around the middle of the last century (*1*). Recently, with the realization of new classes of materials such as graphene (*2*, *3*) and $Z_2$ topological insulators (TIs) (*4*), the interest in SPE has been reignited, attracting a lot of attention from both fundamental and applications perspectives due to the novel phenomena theoretically predicted to be realized based on this effect. For example, proximity-induced superconducting p-n junctions are expected to exhibit superluminescence (*5*) applicable in optoelectronic devices, while heterostructures of TIs and SCs are predicted to harbor topological superconductivity hosting Majorana modes inside vortex cores (*6*).

The quest for achieving topological superconductivity in TI/SC heterostructures has mostly focused on coupling TIs with low-$T_c$ SCs (NbSe$_2$ (*7–11*) and elemental SCs (*12*, *13*)) or with a high-$T_c$ cuprate Bi$_2$Sr$_2$CaCu$_2$O$_{8+x}$ (Bi-2212) (*14–17*). Although TI/NbSe$_2$ platform has led to a widely accepted observation of proximity-induced superconductivity in TIs (*7–11*), the induced superconducting gap of ~1 meV makes experimental characterization of the emergent phase challenging and limits potential future applications of any Majorana modes detected. Optimally-doped cuprate Bi-2212 exhibits a high $T_c$ of ~91 K and a *d*-wave superconducting gap of ~40 meV (*18*), but the detection of the proximity induced superconductivity in the TIs by coupling to it has given conflicting results (*14–17*). This has been attributed to the very short coherence length of Bi-2212 along the c-axis (*14*) and/or the large momentum-space mismatch between the Fermi surface of TI Bi$_2$(Se,Te)$_3$ near the Brillouin zone (BZ) center and that of Bi-2212 near the BZ edge (*14*, *15*). If the hypothesis that Fermi surface matching at the interface actually played an essential role, many Fe-based high-$T_c$ SCs with Fermi surface pockets near the BZ center (*19*) could be a promising alternative for achieving high-$T_c$ SPE in TIs. Moreover, these SCs can achieve extraordinary high superconducting gap (*20*), $T_c$ (*21*) and upper critical fields (*19*). However, in spite of immense potential, there is still no evidence to suggest that Fe-based high-$T_c$ SCs could be used to induce a large gap superconductivity in a proximate non-superconducting TI.

In this work, we report the first observation of superconductivity at the surface of a prototypical topological insulator $Bi_2Te_3$ in proximity to an Fe-based high-$T_c$ SC Fe(Te,Se). To create the desired interface, we deposit a thin film of $Bi_2Te_3$ using molecular-beam epitaxy (MBE) on top of a cleaved $FeTe_{1-x}Se_x$ (x = 0.45, $T_c$ ~ 14.5 K) bulk single crystal (Fig. 1a, Methods). We study $Bi_2Te_3$ films in a range of thicknesses: 1 quintuple layer (QL), 3 QLs and 5 QLs, which span the regimes before and after the Dirac node is formed (22). One of the main challenges in successful $Bi_2Te_3$/Fe(Te,Se) heterostructure growth is the incompatible atomic surface structure of the two materials – Fe(Te,Se) typically cleaves along the (001) orientation to expose a square lattice of Se/Te atoms (Fig. 1d), while $Bi_2Te_3$ has a surface lattice structure with hexagonal symmetry (Fig. 1e). This often results in the formation of two types of $Bi_2Te_3$ film domains rotated in-plane by 30 degrees with respect to one another (Fig. 1b and Fig. S1). However, these domains are large enough that we can easily locate single-domain ~50-100 nm square regions of the sample using scanning tunneling microscopy (STM) (Fig. 2).

To realize topological superconductivity in TI/SC heterostructures, two main conditions need to be satisfied: the existence of helical Dirac states on the surface of a TI and a gap in these states due to induced *s*-wave superconducting pairing. First, we characterize the large-scale electronic structure of our films by using quasiparticle interference (QPI) imaging. This commonly used spectroscopic method relies on elastic scattering of electrons on the surface of a material that can be visualized as "waves" in STM *dI/dV* images (23). For each of the 3 heterostructures studied in this work, we acquire a series of *dI/dV* maps over a large square region of the $Bi_2Te_3$ surface, and examine their Fourier transforms (FTs) as a function of energy (Fig. 2). Based on the thickness-dependent angle-resolved photoemission spectroscopy (ARPES) measurements, $Bi_2Te_3$ films thicker than 2 QLs are expected to be in the regime where the top and the bottom surface do not hybridize (22), and their topological surface state (SS) band structure consists of a single Dirac cone located at Γ. Although backscattering is prohibited in ideal topological insulators, small hexagonal warping of the constant energy contour away from the Dirac point has been shown to open a scattering channel along the Γ-M direction (24–26). On the surface of our $Bi_2Te_3$ films of 3 QL (Fig. 2(e-g)) and 5 QL (Fig. 2(i-k)) thickness, we indeed observe the characteristic angle-dependent QPI pattern with energy-dispersive wave vectors along the Γ-M direction. This is consistent with the previous experiments on cleaved $Bi_2Te_3$ single crystals (*24*, *25*) and films (*26*), and allows us to visualize the topological SS band dispersion (Fig. 2h,l). QPI measurements of 1 QL $Bi_2Te_3$ film also reveal a dispersive QPI pattern in a wide range of energies studied. However, QPI morphology appears to be more isotropic with similar intensity along all angles (Fig. 2c, Fig. S2). This signature is likely a consequence of low spin-momentum locking indicative of degenerate nature of the SS in 1 QL $Bi_2Te_3$ (*9*), which facilitates the observation of the backscattering vector. We note that the observed SS in 1 QL $Bi_2Te_3$ is clearly distinct from the recently reported SS on the surface of Fe(Te,Se) (*27*), which has the Dirac point located 4.4 meV below the Fermi level and ~5 times smaller Fermi wavevector $k_f$. We also emphasize that this is the first QPI imaging of the $Bi_2Te_3$ SS in a range of thicknesses down to the single QL. Importantly, the extracted dispersions match well with the previous ARPES measurements (*9*) (Fig. 2d,h,l), and as such confirm the expected large-scale electronic structure of our $Bi_2Te_3$ films.

We proceed to look for signatures of the induced superconductivity at the surface of $Bi_2Te_3$ and start with the 1QL $Bi_2Te_3$/Fe(Te,Se) heterostructure. This thickness should in principle maximize the magnitude of the induced superconducting gap at the TI surface, which is expected to decrease away from the interface (*28*). Typical *dI/dV* spectrum at the surface of our $Bi_2Te_3$ film reveals a clear energy gap $\Delta_{ind}$ ~ 3.5 meV (Fig. 3a), calculated as one half of the energy difference between the two gap edge peaks. The magnitude of $\Delta_{ind}$ on $Bi_2Te_3$ is spatially homogeneous (Fig. S3) and the same gap has been detected on multiple regions of the sample separated by large distances. It is also significantly larger than that obtained on the surface of non-superconducting, antiferromagnetic parent compound FeTe monolayer grown on $Bi_2Te_3$ reported to be ~1 meV (*29*). The observed gap edge peaks in our *dI/dV* spectra are approximately symmetric with respect to the Fermi level (Fig. 3a), which is consistent with a pairing induced gap. We acquire the average *dI/dV* spectrum as a function of temperature over the identical region of the $Bi_2Te_3$ film and find that the gap persists up to $T_c'$ of 10 K (Fig. 3a). Although $T_c'$ is slightly lower than the bulk $T_c$ ~14.5 K of our Fe(Te,Se) substrate, this difference can be attributed to local $T_c$ variations suspected to occur in Fe(Te,Se) (*30*). As such, these measurements present the first indication that $\Delta_{ind}$ may be a proximity-induced superconducting gap.

To obtain further evidence of the superconducting origin of $\Delta_{ind}$, we investigate the dependence of our *dI/dV* spectra on the applied magnetic field. For a type-II SC such as Fe(Te,Se), magnetic field will penetrate the material in quantized flux quanta below upper critical field $H_{c2}$ in the form of vortices. If the surface of our $Bi_2Te_3$ film becomes superconducting due to proximity to a type-II SC, we would expect to observe Abrikosov vortices as dark regions in a *dI/dV* map acquired near the energy of the superconducting gap (*31*). We acquired a series of *dI/dV* maps at ~$\Delta_{ind}$ (Figs. 3b-g) that each shows a vortex lattice up to the highest magnetic field measured. From these, we calculate the average magnetic flux per vortex to be 2.16 ± 0.07 x $10^{-15}$ $m^2T$ (Figs. S4, S5), in good agreement with the expected single magnetic flux quantum of 2.07 x $10^{-15}$ $m^2T$. The vortices in our $Bi_2Te_3$/Fe(Te,Se) heterostructure persist up to much larger fields compared to $Bi_2Te_3$/$NbSe_2$ (*7*), which is consistent with the extraordinary high $H_{c2}$ of the Fe(Te,Se) family of materials (*32*). The linecut through a vortex core shows the suppression of gap edge peaks inside the vortex (Fig. 3h), also confirming the superconducting origin of $\Delta_{ind}$.

Next, we explore the electronic structure of thicker $Bi_2Te_3$ films, looking for signatures of the induced superconducting gap $\Delta_{ind}$. For each sample, we acquire *dI/dV* spectra under similar conditions, which all show a prominent gap in the density-of-states (Fig. 4a). Similarly to Ref. (*8*), $\Delta_{ind}$ in 1 QL $Bi_2Te_3$ film is approximately equal to that measured on the substrate surface prior to growth (Fig. 4a). In thicker films, $\Delta_{ind}$ gets reduced as the film thickness increases, reaching ~ 1.8 meV for the 3 QL film and ~ 1.5 meV for the 5 QL $Bi_2Te_3$ film, which is consistent with the general trend expected from theoretical calculations (*28*). The measured decay rate of $\Delta_{ind}$ in our heterostructures is also comparable to that in $Bi_2Te_3$/$NbSe_2$ system (*8*) (Fig. 4b). However, we emphasize that $\Delta_{ind}$ in our heterostructures is ~3 times larger than the previous largest $\Delta_{ind}$ (*7–13*), and presents the largest induced $\Delta_{ind}$ to-date in any TI/s-wave SC heterostructure. We also note that none of the *dI/dV* spectra have a flat bottom reaching zero conductance around the Fermi energy. This could be related to localized, incoherent states such as impurity states near the Fermi level. Therefore, simple s-wave BCS function (*33*) cannot reproduce the shape of our spectra (Fig. S6). Although previous experiments on $Bi_2Te_3$/$NbSe_2$ related similar aberration in thicker films to the emergence of topological surface states (*8*), we deem this unlikely as we observe this deviation even in the 1 QL thick $Bi_2Te_3$ film, where Dirac node is yet to be formed.

Finally, we proceed to unveil if emergent superconductivity at the surface of $Bi_2Te_3$ indeed leads to a gap in its SS. To accomplish this, we use QPI imaging to map the SS band crossing the Fermi level in small energy increments. If a gap in the SS exists, the FTs of *dI/dV* maps near the Fermi level should exhibit no detectable QPI signal within the gap. We again start with the 1QL $Bi_2Te_3$/Fe(Te,Se), where the only coherent band expected to cross the Fermi level is the SS band. At energies larger than or comparable to $Δ_{ind}$, Fourier transforms of *dI/dV* maps (Figs. 5a-d) exhibit a QPI pattern qualitatively similar to that observed at higher energies (Fig. 2c). However, as we enter the region within the superconducting gap, we discover a striking suppression of scattering in a range of energies from -2 meV to +2 meV, indicating a full gap in the SS around the Fermi level (Figs. 5e-i). This visual observation is further supported by a more detailed analysis of the QPI intensity as a function of energy, where complete suppression of the QPI pattern occurs exactly within $Δ_{ind}$ and recovers outside of the gap (Fig. 5j). The suppression of QPI has also been confirmed for the 3 QL $Bi_2Te_3$ sample, consistent with a gap in the topological SS (Fig. 5k). Our data presents the first microscopic visualization of a full gap in the coherent SS of $Bi_2Te_3$ due to proximity induced pairing correlations. It also demonstrates that this newly created platform is an emergent, artificially-constructed topological superconductor, with higher $T_c$ and larger $Δ_{ind}$ compared to previous work.

One of the intriguing properties of a topological superconductor in this geometry is that each vortex on its surface is expected to host a Majorana zero mode (*6*). This mode can be detected as a peak in the density-of-states at zero energy. In contrast to bound states in vortices of conventional superconductors (*34*), a Majorana mode should not split in energy away from the vortex center (*10*, *35*). This signature behavior has been observed in >3 QL thick $Bi_2Te_3$ films grown on $NbSe_2$, but is notably absent if the $Bi_2Te_3$ film thickness is reduced (*10*). Interestingly, in our experiments, even in the 1 QL $Bi_2Te_3$ film in which the topological surface state is yet to be formed, we observe a zero-bias peak that gets suppressed, but does not seem to split away from the vortex center (Fig. 3h). Further experiments, using spin-polarized tunneling measurements (*11*), are clearly necessary to shed light on the origin of this peak and how it relates to the recently observed zero bias peak in vortices of pristine FeTe$_{0.55}$Se$_{0.45}$ attributed to a Majorana mode (*36*).

Our results highlight the promise of achieving topological superconductivity with higher $\Delta_{ind}$ and $T_c$ using Fe(Te,Se) and related Fe-based SCs. For example, using a single layer FeSe as a substrate, which exhibits ~5 times larger superconducting gap (*20*) and ~8 times higher $T_c$ (*21*) compared to bulk Fe(Te,Se), could pave the way towards topological superconductivity above liquid nitrogen temperatures. Our experiments could also serve to spark further theoretical work on deeper understanding of the proximity effects in multiband/orbital SCs. Finally, our observations have important implications beyond achieving topological superconductivity with enhanced metrics. Recent experiments and theory suggest that unconventional pairing can be induced at the interface of strong spin-orbit coupled materials (regardless of their topological nature) and SCs by the application of in-plane magnetic field (*37*). Our work introduces a viable platform to explore this pairing tunability without the need of milliKelvin temperatures.

## Methods

High quality FeTe$_{1-x}$Se$_x$ bulk single crystals were grown using self-flux method, and cleaved along the (001) direction in ultra-high vacuum (UHV) to expose a clean surface free of contaminants. Bulk superconducting $T_c$ ~ 14.5 K was confirmed by ex-*situ* magnetization measurements (Fig. S8). Crystalline orientation of the substrate is confirmed by the square lattice observed in STM topographs (Fig. 1d). During the growth process, RHEED pattern (obtained using a 15 keV RHEED gun manufactured by Sentys Inc) was continuously monitored to establish the quality of the substrate and the film surface. To grow Bi$_2$Te$_3$ film, we co-evaporated 99.999% pure Bismuth and 99.99% pure Tellurium from K-cells (Sentys Inc) held at 412°C and 283°C temperatures respectively. The nominal Bi:Te flux ratio used was 1:10, which was determined by using *in-situ* quartz crystal growth monitor (QCM) prior to growth. The film was grown at the rate of 7 minutes per nominally calculated QL of Bi$_2$Te$_3$. Typical post-growth RHEED pattern of our heterostructures exhibits a streaky pattern characteristic of the layer-by layer $Bi_2(Se,Te)_3$ MBE growth (Fig. 1b, Fig. S1). Based on the measured step height of 0.97 ± 0.02 nm between the film (Fig. 1e) and the substrate (Fig. 1d) we can conclude that $Bi_2Te_3$ in Fig. 1c and Fig. 2a-c is exactly 1 QL thick (Fig. 1f). Larger $Bi_2Te_3$ thicknesses have been determined by taking into account the deposition time and the measured Bi flux.

After the growth was completed, the heterostructure was transferred from the MBE to the STM within one hour using a vacuum "suitcase" chamber held at $10^{-11}$ Torr base pressure, which can be directly connected to either MBE or STM chambers. As such, our material is only exposed to UHV conditions during the entire process from the MBE growth to the STM measurement.

STM data was acquired using Unisoku USM1300 STM at the base temperature of 4.5 K (with the exception of temperature dependent data in Fig. 3a). All spectroscopic measurements have been taken using a standard lock-in technique at 915 Hz frequency and bias excitation varying from 0.2 mV to 10 mV as detailed in the figure captions. STM tips used were home-made chemically etched W tips annealed to bright-orange color in UHV. Tip quality has been evaluated on the surface of single crystal Cu(111) prior to performing the measurements presented in this paper. Cu(111) surface was cleaned by repeated cycles of heating and Argon sputtering in UHV before it is inserted into the STM head.

## Acknowledgements

I.Z. gratefully acknowledges the support from Army Research Office Grant Number W911NF-17-1-0399 and National Science Foundation Grant Number NSF-DMR-1654041. The work in Brookhaven was supported by the Office of Science, U.S. Department of Energy under Contract No. DE-SC0012704. J.S. and R.D.Z. were supported by the Center for Emergent Superconductivity, an Energy Frontier Research Center funded by the U.S. Department of Energy, Office of Science. Z.W. acknowledges the support from U.S. Department of Energy, Basic Energy Sciences Grant No. DE-FG02-99ER45747.

**Figures**

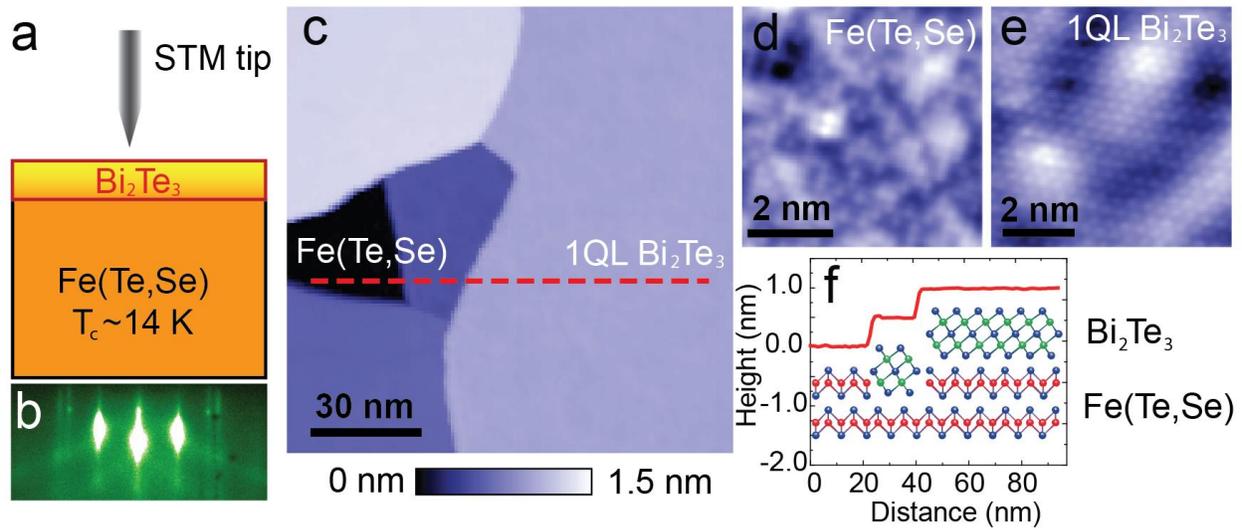

**Figure 1.** (a) Schematic of $Bi_2Te_3$/Fe(Te,Se) heterostructure. STM measurements are probing the exposed top surface of $Bi_2Te_3$. (b) Typical post-growth RHEED image of the MBE-grown heterostructure showing long streaks characteristic of two-dimensional layer-by-layer growth. (c) STM topograph showing (d) the exposed Fe(Te,Se) substrate with square atomic lattice and (e) the $Bi_2Te_3$ film with hexagonal lattice structure. (f) Height profile taken along the dashed red line in (c) consistent with 1 QL of $Bi_2Te_3$ grown on top of Fe(Te,Se). STM setup condition in: (c) $I_{set}$ = 10 pA, $V_{set}$ = -200 mV; (d) $I_{set}$ = 30 pA, $V_{set}$ = -50 mV; (e) $I_{set}$ = 30 pA, $V_{set}$ = -50 mV.

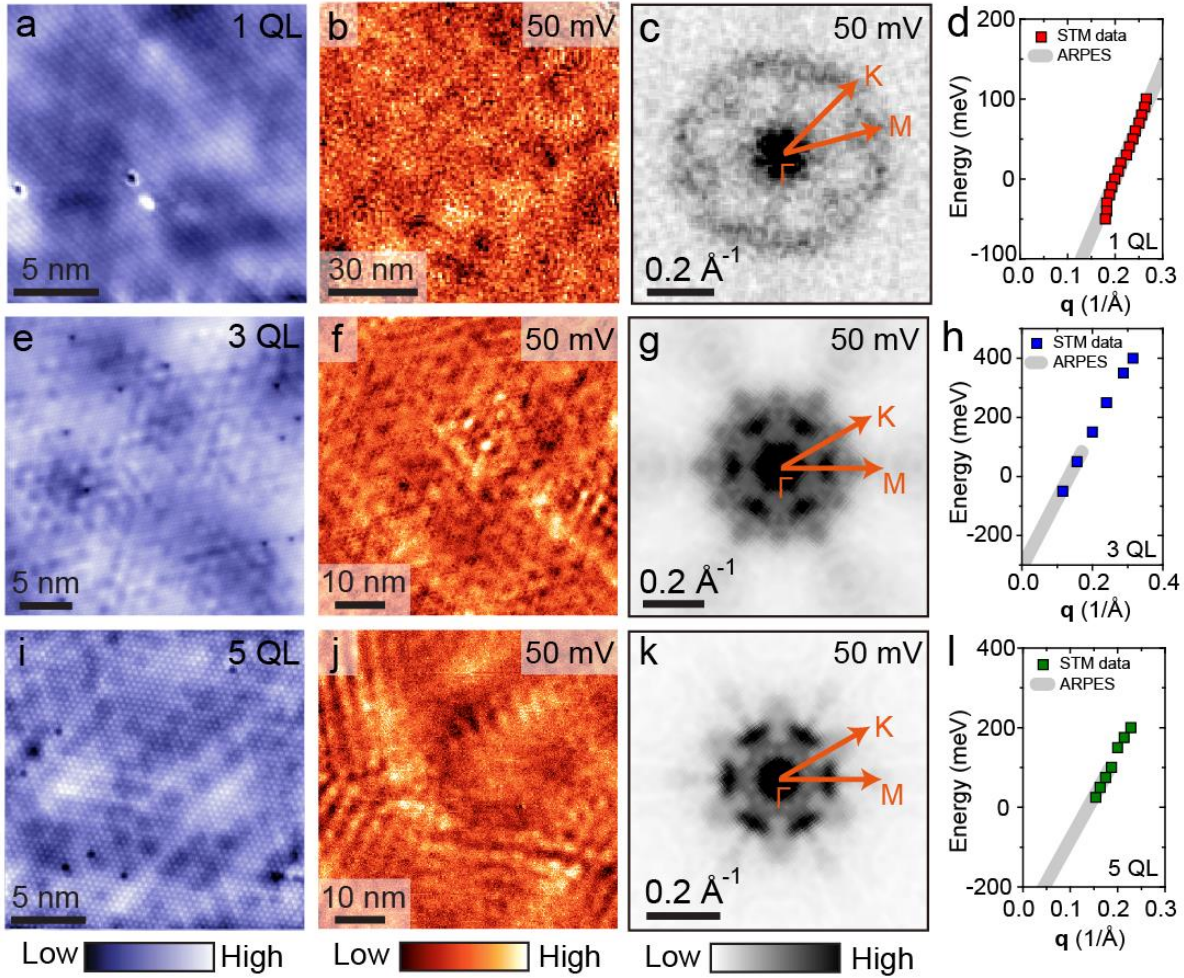

**Figure 2.** Thickness-dependent quasiparticle interference imaging of the $Bi_2Te_3$ surface states. (a,e,i) Typical STM topographs, (b,f,j) large region $dI/dV$ maps acquired at 50 mV, (c,g,k) their associated Fourier transforms and (d,h,l) extracted QPI dispersion along Γ-M direction for 1 QL, 3 QL and 5 QL thick $Bi_2Te_3$ films, respectively. Thick gray lines in (d,h,l) represent fit to our data based on ARPES measurements in Ref. (*22*), shifted upwards by (d) ~150 meV, (h) ~100 meV and (l) ~100 meV to match the QPI dispersions. This rigid band shift can be easily explained by the difference in substrate choice and possibly different impurity concentration, as the shift of the same magnitude is also observed between ARPES studies of for example 1 QL $Bi_2Te_3$/$NbSe_2$ (*9*) and 1QL $Bi_2Te_3$/Si (*22*). The FTs in (g,k) have been 6-fold symmetrized to increase the signal-to-noise. STM setup condition in: (a) $I_{set}$ = 80 pA, $V_{set}$ = -40 mV; (b) $I_{set}$ = 450 pA, $V_{set}$ = -150 mV, $V_{exc}$ = 10 mV (zero-to-peak); (e) $I_{set}$ = 30 pA, $V_{set}$ = 6 mV; (f) $I_{set}$ = 100 pA, $V_{set}$ = 50 mV, $V_{exc}$ = 10 mV; (i) $I_{set}$ = 100 pA, $V_{set}$ = 10 mV; (j) $I_{set}$ = 50 pA, $V_{set}$ = 50 mV, $V_{exc}$ = 10 mV.

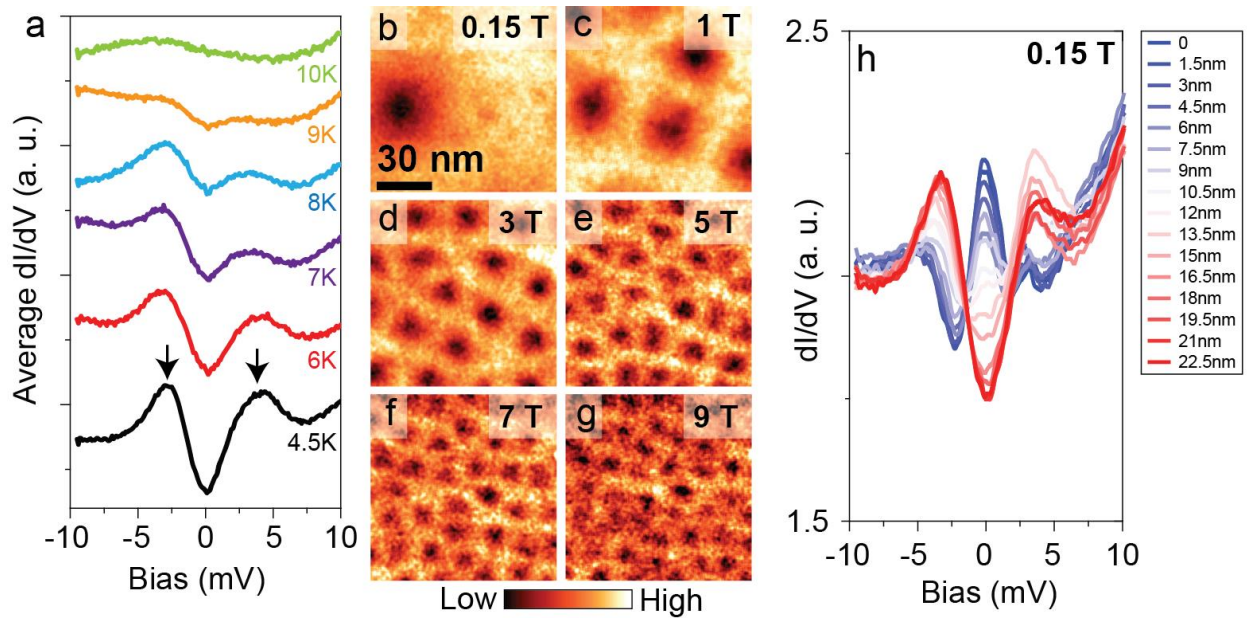

**Figure 3.** Temperature dependence of *dI/dV* spectra and Abrikosov vortex lattice imaging. (a) Average *dI/dV* spectrum on 1 QL $Bi_2Te_3$ as a function of temperature, vertically offset for clarity, showing the gap closing at ~10 K. All spectra have been acquired under the same conditions over the identical 20 nm region of the sample. (b-g) *dI/dV* maps acquired at -3 mV bias over the region of the sample shown in Fig. 2b in a range of magnetic fields from 0.15 T to 9 T. Each dark region in (b-g) represents a single vortex, and the vortex lattice is visible up to the highest field studied. (h) Radially averaged *dI/dV* spectra as a function of distance from the center of the vortex (0.15 T in (b)). Zero bias peak and the suppression of coherence peaks are apparent in the vortex center (darkest blue curve). STM setup condition in: (a) $I_{set}$ = 60 pA, $V_{set}$ = -10 mV, $V_{exc}$ = 0.2 mV; (b-g) $I_{set}$ = 3 pA, $V_{set}$ = -3 mV, $V_{exc}$ = 1.5 mV (zero-to-peak); (h) $I_{set}$ = 60 pA, $V_{set}$ = -10 mV, $V_{exc}$ = 0.45 mV.

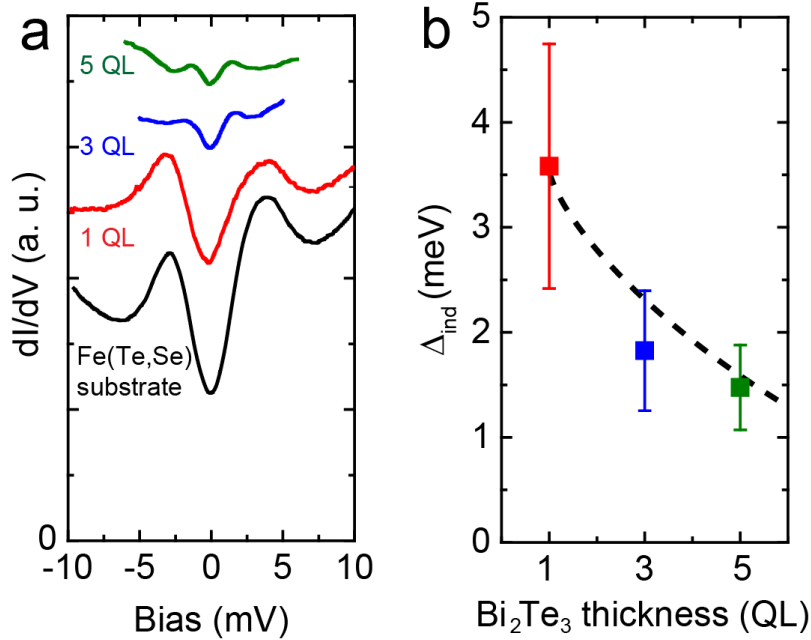

**Figure 4.** Evolution of the induced superconducting gap $\Delta_{ind}$ with $Bi_2Te_3$ film thickness. (a) Average *dI/dV* spectra obtained on: Fe(Te,Se) substrate (black), 1 QL (red), 3 QL (blue) and 5 QL (green) $Bi_2Te_3$ film grown on superconducting Fe(Te,Se). (b) $\Delta_{ind}$ as a function of $Bi_2Te_3$ film thickness. Dashed line in (b) represents the decay rate of the $\Delta_{ind}$ on the surface of $Bi_2Te_3$ induced by $NbSe_2$ from Ref. (*8*), scaled by a multiplicative factor to match the $\Delta_{ind}$ for 1 QL in our case. STM setup condition in: (a) (black) $I_{set}$ = 30 pA, $V_{set}$ = 10 mV, $V_{exc}$ = 0.5 mV (zero-to-peak); (red) same as Fig. 3a; (blue) $I_{set}$ = 30 pA, $V_{set}$ = 5 mV, $V_{exc}$ = 0.4 mV (zero-to-peak); (green) $I_{set}$ = 30 pA, $V_{set}$ = 6 mV, $V_{exc}$ = 0.4 mV (zero-to-peak).

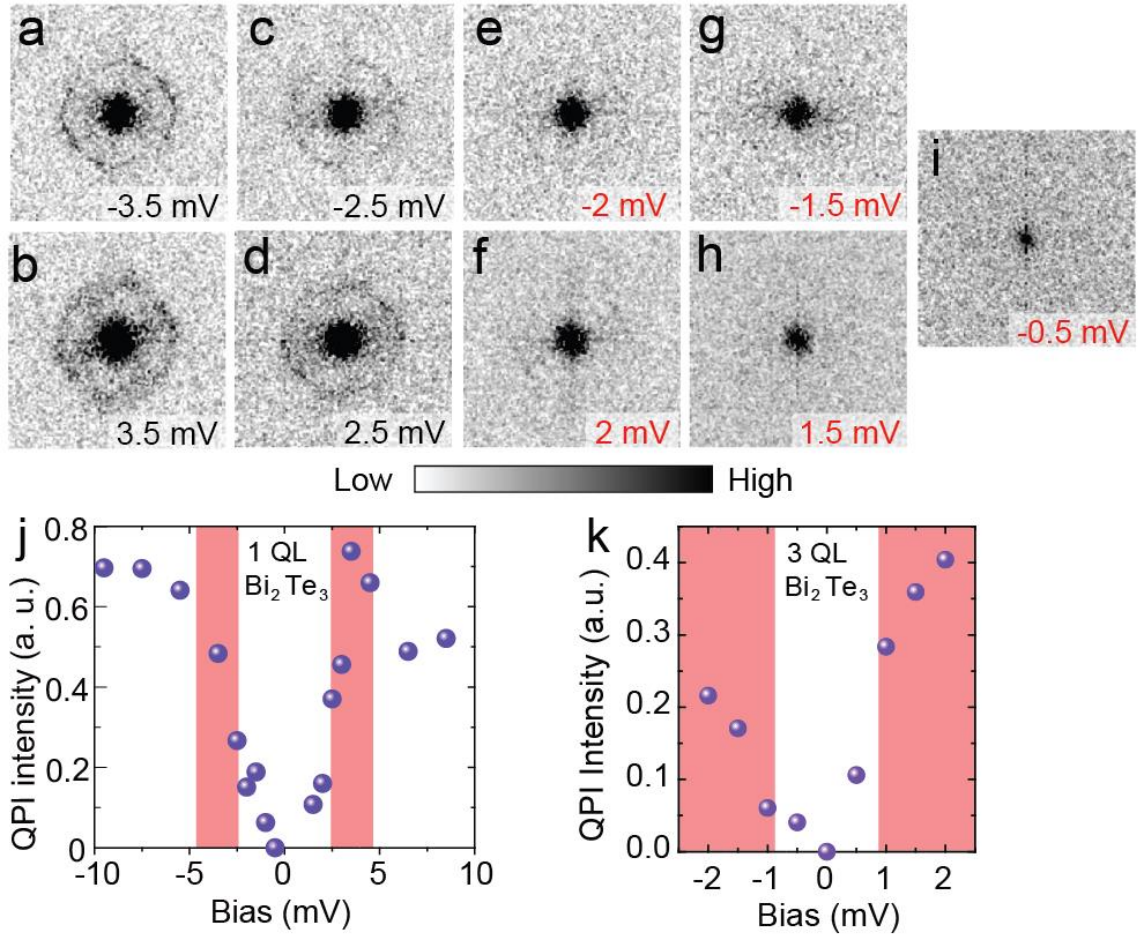

**Figure 5.** Suppression of QPI within the superconducting gap. (a-i) Fourier transforms of *dI/dV* maps acquired over the region in Fig. 2b. While distinct quasi-circular QPI peak can be seen in (a-d), the same peak is obviously suppressed in (e-i). Normalized QPI intensity relative to the background as a function of STM bias for (j) 1 QL thick and (k) 3 QL thick $Bi_2Te_3$ films (Fig. S7). Position of the pink lines in (j,k) indicate the approximate energy of the observed gap peaks, while the width of the lines corresponds to thermal broadening expected at the measurement temperature. All data in (a-i) have been acquired under the same experimental conditions (0.33 GΩ junction resistance with 1 mV zero-to-peak lock-in excitation).